\documentclass{webofc}
\usepackage[varg]{txfonts}   
\pdfoutput=1
\usepackage{pstricks}
\usepackage{color}
\usepackage{amssymb,amsmath,bbm}
\usepackage{epsf}
\usepackage{epsfig}
\usepackage{afterpage}
\usepackage{longtable}
\usepackage{latexsym,mathrsfs,dsfont}
\usepackage{graphics}
\usepackage{url}
\usepackage{paralist}
\usepackage{bbold}

\newcommand{\gev}{\, {\rm GeV}}
\newcommand{\mev}{\, {\rm MeV}}

\newcommand{\ord}{\mathcal{O}}
\newcommand{\bsi}{B_6^{(1/2)}}
\newcommand{\bei}{B_8^{(3/2)}}

\def\epe{\varepsilon'/\varepsilon}
\newcommand{\be}{\begin{equation}}
\newcommand{\ee}{\end{equation}}
\newcommand{\bi}{\begin{itemize}}
\newcommand{\ei}{\end{itemize}}

\newcommand{\muEW}{{\mu_\mathrm{ew}}}
\newcommand{\muLow}{{\mu}}

\def\kpn{K^+\rightarrow\pi^+\nu\bar\nu}

\def\klpn{K_{L}\rightarrow\pi^0\nu\bar\nu}
\def\klpll{K_{L}\rightarrow\pi^0 l^+l^-}

\usepackage{tikz}

\woctitle{QCD@Work 2018}

\begin{document}

\begin{flushright}
        {AJB-18-8}
\end{flushright}

\vspace{7mm}

\title{\boldmath The Dual QCD @ Work: 2018 \unboldmath}
%
%
\author{ Andrzej~J.~Buras\fnsep\thanks{Talk given at QCD@Work 2018, Matera, June 2018.} 
}
\institute{TUM Institute for Advanced Study, Lichtenbergstr. 2a, D-85748 Garching, Germany}

\abstract{%
The Dual QCD (DQCD) framework, based on the ideas of 't Hooft and Witten, and developed  
by Bill Bardeen, Jean-Marc G{\'e}rard and myself in the 1980s is not QCD, a theory of quarks and gluons, but a successful low energy approximation of it when 
applied to $K\to\pi\pi$ decays and $K^0-\bar K^0$ mixing.                        After years of silence, starting with 2014, this framework has been further 
developed in order to improve the SM prediction for the ratio $\epe$, the $\Delta I=1/2$ rule and $\hat B_K$. Most importantly, this year it has been used for the calculation of {\em all}  $K\to\pi\pi$ hadronic matrix elements of BSM operators which opened the road for the general
study of $\epe$ in the context of the SM effective theory (SMEFT). This talk 
summarizes briefly the past successes of this framework and discusses recent 
developments which lead to a master formula for $\epe$ valid in {\em any} extension of the SM. This formula should facilitate the search for new physics responsible 
for the $\epe$ anomaly hinted by 2015  results from 
lattice QCD and DQCD. }
\maketitle
\section{Introduction}
\label{intro}
Among the main stars of kaon flavour physics is the ratio $\epe$ that describes 
the amount of direct CP violation in $K_L\to\pi\pi$ decays relative to the
indirect one. The latter one is related to  $K^0-\bar K^0$ mixing, where also 
the $K_L-K_S$ mass difference, $\Delta M_K$, is an important quantity. Among 
rare kaon decays the most important roles these days are played by  rare decays $\kpn$ and $\klpn$, soon to be measured by NA62 and KOTO experiments, respectively. 

Various aspects of these decays and observables are reviewed at length in 
\cite{Buras:2013ooa,Buras:2018wmb}. Here we will concentrate our discussion 
on $\epe$ that has been measured already many years ago with
the experimental world
average from NA48 \cite{Batley:2002gn} and KTeV \cite{AlaviHarati:2002ye, 
Abouzaid:2010ny} collaborations given by
\begin{align}
  \label{eq:epe:EXP}
  (\epe)_\text{exp} & = (16.6 \pm 2.3) \times 10^{-4}\,.
\end{align}

There are at least two reasons for concentrating on this ratio. First
the situation of $\epe$ in the SM can be briefly summarized as follows:
\begin{itemize}
\item
The analysis of $\epe$ by the RBC-UKQCD collaboration based on their lattice QCD
calculation of $K\to \pi\pi$ matrix elements \cite{Bai:2015nea, Blum:2015ywa}, as well as the analyses performed in  \cite{Buras:2015yba,Kitahara:2016nld}
that are based on the same
matrix elements but also include isospin breaking effects,
find $\epe$ in the ballpark of $(1-2) \times 10^{-4}$. This is
by one order of magnitude below the data, but with an error in the ballpark of
$5\times 10^{-4}$. Consequently, based on these analyses, one can talk about an $\epe$ anomaly of at most~$3\sigma$.
\item
An independent analysis based on hadronic matrix elements from the Dual QCD (DQCD)
approach \cite{Buras:2015xba, Buras:2016fys} gives a strong support to these values
and moreover provides an \textit{upper bound} on $\epe$ in the ballpark of $6\times
10^{-4}$.
\item
A different view has been expressed in \cite{Gisbert:2017vvj} where, using ideas
from chiral perturbation theory, the authors find $\epe = (15 \pm 7) \times 10^{-4}$.
While in
agreement with the measurement, the large uncertainty, that expresses
the difficulties in matching long distance and short distance contributions
in this framework, does not allow for clear-cut conclusions.
Consequently, values above $2\times 10^{-3}$, that are rather unrealistic from the
point of view of lattice QCD and DQCD, are not excluded in this approach.
\end{itemize}

Based on the results from RBC-UKQCD and the DQCD approach of 2015 
a number of analyses have been performed in specific models beyond the SM (BSM)
with the goal to obtain a  sufficient upward shift in $\epe$ and thereby
its  experimental value. They are collected in Table~1
of \cite{Buras:2018wmb}. 
An important limitation of these analyses is that they address the $\epe$
anomaly only in models in which new physics (NP) enters exclusively through
modifications of the Wilson coefficients of SM operators. However, generally,
BSM operators with different Dirac structures -- like the ones resulting from
tree-level scalar exchanges and leading to scalar and tensor operators -- or
chromo-magnetic dipole operators could play a significant role in $\epe$.
Until recently, no quantitative judgment of the importance of such operators
was possible because of the absence of even approximate calculations of the
relevant hadronic matrix elements in QCD. 

This brings me to the second reason for concentrating here on $\epe$. 
Namely, recently  a significant theoretical  progress has been made  towards the general search for  NP responsible for the hinted $\epe$ anomaly. This progress is based on the following 
analyses:
\begin{itemize}
\item
the first to date calculations of the $K\to\pi$ and $K\to\pi\pi$ matrix elements of 
the chromo-magnetic dipole operators by lattice QCD
\cite{Constantinou:2017sgv} and DQCD \cite{Buras:2018evv}, respectively and in particular
the calculation of $K\to\pi\pi$ matrix elements of {\em all} four-quark BSM operators,
including scalar and tensor operators, by DQCD \cite{Aebischer:2018rrz}.
\item
derivation of a master formula for $\epe$ \cite{Aebischer:2018quc}, which
can be applied to any theory beyond the SM in which the
Wilson coefficients of all contributing operators have been calculated at the
electroweak scale. The relevant hadronic matrix elements of BSM operators used 
in this formula are
from the DQCD, as lattice QCD did not calculate them yet, 
and the SM ones from lattice QCD.
\item
the first to date  model-independent anatomy of the ratio $\epe$
in the context of  the $\Delta S = 1$ effective theory with operators invariant
under QCD and QED and in the context of the Standard Model Effective Field Theory (SMEFT) with the operators invariant under the full SM gauge group \cite{Aebischer:2018csl}.
\end{itemize}

The main goal of this talk is to review briefly these advances and to stress 
that they would not be possible in this decade without the existence of 
DQCD approach \cite{Bardeen:1986vz, Buras:1985yx, Buras:2014maa, Buras:2015xba}.

We begin our presentation in Section~\ref{sec:2} by recalling the general structure of the effective Hamiltonian valid in any BSM theory and summarizing the number of BSM operators one has to consider. In Section~\ref{sec:3} we will
briefly recall the main virtues of the DQCD approach.  In this section we will stress the importance of the so-called {\em meson evolution}, an important ingredient of DQCD, which describes {\em analytically} very important dynamics at scales lower than $1\gev$. 
  In Section~\ref{sec:4} we will briefly review the present status of 
$\epe$ anomaly and in Section~\ref{sec:5} we will present 
 a master formula for $\epe$ that is valid in 
any theory beyond the SM \cite{Aebischer:2018quc}.  In Section~\ref{sec:6} 
we summarize the main lessons from the EFT and SMEFT analyses of $\epe$ in 
\cite{Aebischer:2018csl}. We conclude in Section~\ref{sec:7} with an 
outlook for coming years.

\section{Effective Hamiltonian for $K\to\pi\pi$ Beyond the SM}\label{sec:2}
The technology 
of operator product expansion and renormalization group can easily be extended to any model containing 
heavy particles and new interactions. Starting with a NP scale 
$\Lambda_\text{NP}$ which could still be of order of a few TeV but also much larger, like 
100 TeV, one can consecutively integrate out heavy particles and 
construct a series of effective theories until one reaches the electroweak 
scale. Below it only gluons, photon and SM quarks and leptons except for the top quark appear 
as dynamical degrees of freedom. Subsequently the known RG evolution down to 
scales relevant for a given decay is performed. The resulting effective Hamiltonian at these low energy scales has the following general structure:
\be
{\cal H}_\text{eff}= \sum_i C_i \mathcal{O}_i^\text{SM}+\sum_j C_j^\text{NP} \mathcal{O}_j^\text{NP}\,.
\ee
The four objects appearing in this formula are as follows:
\begin{itemize}
\item
$\mathcal{O}_i^\text{SM}$ are local operators present in the SM 
and $\mathcal{O}_j^\text{NP}$ are new operators having typically new Dirac structures, in particular scalar-scalar and tensor-tensor operators. They will play an important role soon.
\item
$C_i$ and   $C_j^\text{NP}$ are the Wilson coefficients of these operators. But whereas  $C_j^\text{NP}$  are only non-vanishing in the presence of NP, 
\be
C_i=C_i^\text{SM}+\Delta_i^\text{NP}.
\ee
This means that in the presence of NP the Wilson coefficients of SM operators 
are generally modified.
\end{itemize}

The amplitudes for $K\to\pi\pi$ decays 
can now be written as follows
\be\label{AKpipi}
{\cal A}(K\to \pi\pi)= \sum_i C_i(\mu)\langle \pi\pi| \mathcal{O}_i^\text{SM}(\mu)|K\rangle
+\sum_j C_j^\text{NP}(\mu) \langle \pi\pi| \mathcal{O}_j^\text{NP}(\mu)|K\rangle\,.
\ee

The coefficients $C_i$ and $C_j^\text{NP}$ can be calculated in the RG improved perturbation theory, although generally the analysis is rather complicated because of the presence of many operators  that mix under renormalization. 
As far as QCD and QED corrections are concerned, they have been known
already for 25~years at NLO \cite{Buras:1991jm, Buras:1992tc, Buras:1992zv,
Ciuchini:1992tj, Buras:1993dy, Ciuchini:1993vr} and 
for the BSM operators two-loop anomalous
dimensions have been known \cite{Ciuchini:1997bw, Buras:2000if} for almost two
decades. First steps towards NNLO predictions
for $\epe$  have been made in \cite{Bobeth:1999mk, Buras:1999st, Gorbahn:2004my,
Brod:2010mj} and further progress towards a complete NNLO result is under
way~\cite{Cerda-Sevilla:2016yzo}. 

In the 
context of  the so-called SM effective theory not only QCD and QED effects below the electroweak scale but also 
electroweak effects above this scale and in particular the top quark Yukawa coupling have an important impact on the values of the coefficients  $C_i$ and $C_j^\text{NP}$  at the electroweak scale. But fortunately all these effects are already 
summarized for EFT in \cite{Aebischer:2015fzz,Jenkins:2017jig} and for 
SMEFT in  \cite{Jenkins:2013zja,Jenkins:2013wua,Alonso:2013hga}.

Until recently only the first sum in (\ref{AKpipi}) could be evaluated because
the matrix elements of the new operators were unknown. The new development 
is the calculation of the matrix elements of all BSM four-quark operators
 using  DQCD in the SU(3) chiral limit  \cite{Aebischer:2018rrz}. As it will still take some time before corresponding results in lattice QCD will be available, already these approximate results from DQCD can teach us a lot about the relevance of various operators. Putting aside the chromomagnetic dipole operator, for a given chirality
there are 13 BSM four-quark operators in the EFT and 7 BSM four-quark operators
 in the SMEFT for which hadronic matrix elements have to be calculated. 
This counting is demonstrated in  \cite{Aebischer:2018rrz} for the case 
of $(u,d,s)$ quarks. The full list of operators including those with  $(c,b)$ quarks can be found in Table 1 of  \cite{Aebischer:2018quc}.

\section{Basics of Dual QCD Approach}\label{sec:3}
\subsection{Basic framework}
This analytic approach to $K\to\pi\pi$ decays and $K^0-\bar K^0$ mixing originated in the ideas of 't Hooft and Witten
\cite{'tHooft:1973jz,'tHooft:1974hx,Witten:1979kh,Treiman:1986ep} based 
on large number $N$ of colours. In this limit  QCD at very low momenta becomes a  theory of weakly interacting mesons with the coupling $\ord(1/N)$ and in particular in the strict large $N$ limit becomes a free theory 
of mesons simplifying the calculations significantly. With  non-interacting mesons the factorization of matrix elements 
of four-quark operators  into matrix elements of quark currents and quark 
densities  used adhoc in the 1970s and early 1980s is automatic and can be considered as a property of QCD in this limit. But factorization cannot be the whole 
story as the most important QCD effects related to asymptotic freedom 
are related to non-factorizable contributions generated by exchanges of gluons.
In DQCD this role is played by meson loops that represent 
dominant non-factorizable contributions at the very low energy scale. Calculating these loops one discovers 
then  that factorization in question 
does not take place at values of $\mu \ge 1\gev$ at which Wilson coefficients 
are calculated but  rather at very low  momentum transfer between colour-singlet currents or densities.

The explicit calculation of the contributions of pseudoscalars to hadronic matrix elements of local operators is based on  a truncated chiral Lagrangian describing the low energy 
interactions of the lightest mesons \cite{Chivukula:1986du,Bardeen:1986vp,Bardeen:1986uz}
\be\label{chL}
L_{tr}=\frac{F^2}{8}\left[\text{Tr}(D^\mu UD_\mu U^\dagger)+r\text{Tr}(mU^\dagger+\text{h.c.})-\frac{r}{\Lambda^2_\chi}\text{Tr}(mD^2U^\dagger+\text{h.c.})\right]
\ee
where 
\be\label{UU}
U=\exp(i\sqrt{2}\frac{\Pi}{F}), \qquad 
\Pi=\sum_{\alpha=1}^8\lambda_\alpha\pi^\alpha
\ee
is the unitary chiral matrix describing the octet of light pseudoscalars. 
The parameter $F$ is related to  the weak decay constants $F_\pi\approx 130\mev$ 
and $F_K\approx 156\mev$ through
\be\label{FpiFK}
F_\pi=F\left(1+\frac{m_\pi^2}{\Lambda^2_\chi}\right), \qquad F_K=F\left(1+\frac{m_K^2}{\Lambda^2_\chi}\right),
\ee
so that $\Lambda_\chi\approx 1.1\gev$.
The diagonal mass matrix $m$ involving $m_u$, $m_d$ and $m_s$ is such
that 
\be\label{rr}
r(\mu)=\frac{2 m_K^2}{m_s(\mu)+m_d(\mu)},
\ee
with $r(1\gev)\approx 3.75\gev$ for $(m_s+m_d)(1\gev)\approx 132\mev$. 

In order to calculate the matrix elements of the local operators in question we 
need meson representations of colour-singlet quark currents and densities. 
They are directly obtained from the effective Lagrangian in (\ref{chL}) and 
are given respectively as follows
\be\label{VAc}
\bar q^b_L\gamma_\mu q^a_L=i\frac{F^2}{8}\left\{(\partial_\mu U)U^\dagger-U(\partial_\mu U^\dagger)+
\frac{r}{\Lambda^2_\chi}\left[(\partial_\mu U)m^\dagger-m(\partial_\mu U^\dagger)\right]\right\}^{ab},
\ee
\be\label{RLd}
\bar q_R^b q_L^a=-\frac{F^2}{8}r\left[U-\frac{1}{\Lambda_\chi^2}\partial^2U\right]^{ab}\,,
\ee
with $U$ turned into $U^\dagger$ under parity.

At the tree level, corresponding to leading order in $1/N$, one uses these representations  to simply express the operators in terms of the meson fields and expands the matrix $U$ in powers of $1/F$. 
For $K\to\pi\pi$ decays the relevant contribution to hadronic matrix elements is read off from terms involving only one kaon  and two pions. 
However, in the large $N$ limit the hadronic matrix elements factorize and 
in order to combine them with the Wilson coefficients loops in the meson 
theory have to be calculated. In contrast to chiral perturbation theory, in 
DQCD a physical cut-off $\Lambda$ is used in the integration over loop momenta.
 As discussed in detail in \cite{Bardeen:1986vz,Buras:2014maa} this allows 
to achieve  much better matching with short distance contributions than it
is possible in chiral perturbation theory which uses dimensional regularization.  The cut-off $\Lambda$ is typically chosen around $0.7\gev$ when only pseudoscalar mesons are exchanged in the
loops \cite{Bardeen:1986vz} and can be increased to $0.9\gev$ when contributions from lowest lying vector mesons are taken into account as done in  \cite{Buras:2014maa}. These calculations are done in a momentum scheme but as 
demonstrated in \cite{Buras:2014maa} they can be matched to the commonly 
used NDR scheme. Once this is done it is justified to set $\Lambda\approx \mu$.
We ask sceptical readers to study a detailed exposition of DQCD in \cite{Buras:2014maa} where also the differences from the usual chiral perturbation calculations are emphasized.

\subsection{Grand view of DQCD} 
The application of DQCD to weak decays consists in any model of the following 
steps:

{\bf Step 1:} At $\Lambda_\text{NP}$ one integrates out the heavy degrees of freedom and performs the RG evolution including Yukawa couplings and all gauge interactions present in the SM down to the electroweak scale. This evolution involves in addition to SM 
operators also BSM operators. This is SMEFT.

{\bf Step 2:} At the electroweak scale $W$, $Z$, top quark and the Higgs are integrated out and the SMEFT is matched on to EFT with only SM quarks but top-quark, the photon and the gluons. Subsequently QCD and QED evolution is performed down to scales $\ord(1\gev)$.

{\bf Step 3:} Around scales $\ord(1\gev)$ the matching to the theory of mesons
is performed and meson evolution to the factorization scale is performed.

{\bf Step 4:} The matrix elements of all operators are calculated in the 
large $N$ limit, that is using factorization of matrix elements into products of currents or densities.

We do not claim that these are all QCD effects responsible for non-leptonic 
transitions but these evolutions based entirely on non-factorizable QCD
effects both at short distance and long distance scales appear to be the main 
bulk of QCD dynamics responsible for the $\Delta I=1/2$ rule, $\epe$ and $K^0-\bar K^0$ mixing.
\subsection{Past successes of DQCD}
Past successes of this approach have been recently reviewed in 
\cite{Buras:2018wmb}. In particular the non-perturbative parameter $\hat B_K$ 
has been predicted already in 1987 to be close to its large $N$ value $0.75$ 
and found in 2014 with higher precision to be  $\hat B_K=0.73\pm0.02$ \cite{Buras:2014maa} in a very good agreement with the present FLAG average
$\hat B_K=0.762\pm 0.010$ \cite{Aoki:2016frl}. 

DQCD also allowed for the first time to identify already in 1986 the dominant 
mechanism behind the $\Delta I=1/2$ rule \cite{Bardeen:1986vz}.
In the framework of DQCD 
the current-current operators and not QCD penguins  are responsible dominantly for this rule.  In fact this dynamics is rather simple and follows the general 
pattern outlined above. It is just short distance 
(quark-gluon) evolution of current-current operators down to scales  $\ord(1\gev)$ 
followed by meson evolution down to scales $\ord(m_\pi)$ at which the hadronic 
matrix elements factorize and can easily be calculated. Improved calculations 
in 2014 \cite{Buras:2014maa} resulted in 
\be\label{DRULEL}
\left(\frac{{\rm Re}A_0}{{\rm Re}A_2}\right)_{{\rm DQCD}}=16.0\pm 1.5, \qquad
\left(\frac{{\rm Re}A_0}{{\rm Re}A_2}\right)_{{\rm lattice~QCD}}=31.0\pm11.1 \,,
\ee
where we also give the result from the RBC-UKQCD collaboration \cite{Bai:2015nea}. Also lattice result is governed by current-current operators. But the uncertainty is still very large and it 
will be interesting to see whether the lattice QCD will be able to come closer to the
data 
\be\label{DRU}
\left(\frac{{\rm Re}A_0}{{\rm Re}A_2}\right)_{{\rm exp}}=22.35\,,
\ee
than it is possible using DQCD.

I doubt that 
the remaining piece can be fully explained by NP as this would lead to
a large fine-tuning in $\Delta M_K$ as demonstrated in \cite{Buras:2014sba}. But  as analysed in that paper a colour octet of heavy gluons ($G^\prime$) could
bring the theory closer to the data.  It is however likely that also final state interactions (FSI) as stressed by Pich and collaborators and additional subleading corrections not included in the DQCD result in (\ref{DRULEL}) could be responsible for
 the missing piece. Yet, I do not think that the present analytic 
methods like DQCD or the methods advocated by Pich and collaborators, as reviewed recently in \cite{Gisbert:2017vvj}, are sufficiently powerful 
to answer the question at which level NP enters the amplitudes 
${\rm Re}A_0$ and ${\rm Re}A_2$.  Here lattice QCD should provide valuable
answers and I am looking forward to improved results on these two amplitudes 
from RBC-UKQCD collaboration and other lattice groups. This would 
provide two additional important constraints on NP models. 

DQCD is really not QCD, the theory of quarks and gluons, but as we have seen 
above and we will see below a successful low energy approximation of QCD.
Moreover, it has several virtues:
\begin{itemize}
\item
It is an efficient approximate method for obtaining results for non-leptonic 
decays, years and even decades before useful results from numerically sophisticated and demanding lattice calculations could be obtained.
\item
It is the only existing method allowing to study analytically the dominant 
dynamics below $1\gev$ scale, represented by the meson evolution, which 
turns out to have the pattern of operator mixing, both for SM and BSM operators,
 to agree with the one found perturbatively at short distance scales. This allows for satisfactory, even if approximate, matching between Wilson coefficients and hadronic matrix elements.
\item
It provides insight in the purely numerical results obtained by lattice QCD   for both $K^0-\bar K^0$ hadronic matrix elements of 
BSM operators  and of SM QCD and electroweak penguin operators relevant for $\epe$. See below.
\item
Most importantly it allowed already now to evaluate $K\to\pi\pi$ matrix elements of all BSM operators providing in this manner for the first time a global view
on NP contributions to $\epe$.
\end{itemize}

\section{More on $\epe$ Anomaly}\label{sec:4}
  While the results based on the hadronic matrix elements from RBC-UKQCD 
lattice collaboration, suggest some evidence for the presence of NP in hadronic $K$ decays,
the large uncertainties in the hadronic matrix elements in question do not yet
preclude that eventually the SM will agree with data. In this context the 
upper bounds on the matrix elements of the
dominant penguin operators from DQCD \cite{Buras:2015xba} are important as they give presently the strongest support to the anomaly 
in question, certainly stronger than present lattice results. To see this in 
explicit terms  let us look at the parameters $\bsi$ and $\bei$ 
that represent the relevant hadronic matrix elements of the QCD penguin operator $Q_6$ 
and the electroweak penguin operator $Q_8$, respectively. 

In the strict large $N$ limit \cite{Buras:1985yx,Bardeen:1986vp,Buras:1987wc} one simply has $\bsi=\bei=1$, but RBC-UKQCD results \cite{Bai:2015nea,Blum:2015ywa} imply
\cite{Buras:2015yba,Buras:2015qea}
\be\label{Lbsi}
\bsi=0.57\pm 0.19\,, \qquad \bei= 0.76\pm 0.05\,, \qquad (\mbox{RBC-UKQCD}),
\ee
and this suppression of both parameters below unity, in particular of $\bsi$, 
is the main origin of the strong suppression of $\epe$ within the SM 
below the data. Yet in view of the large error in $\bsi$ one could 
be sceptical about claims made by me and my collaborators that there is 
NP in $\epe$. Future lattice results could in principle raise $\bsi$ towards its large $N$ value and above $\bei$ bringing the SM result for $\epe$ close
to its experimental value. 

However, the analyses of $\bsi$ and $\bei$ within DQCD in
\cite{Buras:2015xba,Buras:2016fys} show that such a situation is rather 
unlikely. Indeed, in this approach going beyond the strict large $N$ limit  
one can understand the suppression of $\bsi$ and $\bei$ below the unity 
 as the effect of the meson 
evolution from scales  $\mu=\ord(m_\pi)$  at which $\bsi=\bei=1$ is valid 
to  $\mu=\ord(1\gev)$ at which Wilson coefficients of $Q_6$ and $Q_8$ are 
evaluated \cite{Buras:2015xba}. This evolution has to be matched to the usual perturbative quark evolution for scales higher than $1\gev$ and in fact the supressions in question
and the property that $\bsi$ is more strongly suppressed than $\bei$ are 
consistent with the perturbative evolution of these parameters above  
$\mu=\ord(1\gev)$. Thus we are rather confident that \cite{Buras:2015xba}
\be\label{NBOUND}
\bsi< \bei < 1 \, \qquad ({\rm dual~QCD}).
\ee
Explicit calculation in this approach gives $B_8^{(3/2)}(m_c)=0.80\pm 0.10$.
The result for $\bsi$ is less precise but  in agreement with
(\ref{Lbsi}). For further details, see \cite{Buras:2015xba}.

It should be recalled that in the past values $\bsi=\bei=1.0$ 
have been combined in phenomenological applications with the Wilson coefficients evaluated at scales $\mu=\ord(1\gev)$. The discussion above shows that this 
is incorrect. The meson evolution from  $\mu=\ord(m_\pi)$  to $\mu=\ord(1\gev)$ 
has to be performed and this effect turns out to be stronger than the 
scale dependence of $\bsi$ and $\bei$ in the perturbative regime, where it is
 very weak.

Additional support for the small value of $\epe$ in the SM comes from the  reconsideration of the role of final state interactions (FSI) in $\epe$ in
\cite{Buras:2016fys}. In this paper using the DQCD approach 
we have demonstrated that in contrast to claims made by 
chiral perturbation theory practitioners, in particular Pich and collaborators 
as summarised recently in \cite{Gisbert:2017vvj}, FSI are not expected to be
important for $\epe$. But we agree with these authors that FSI 
are likely to be important for the $\Delta I=1/2$  rule. I should remark
that in private conversations the authors of \cite{Gisbert:2017vvj} 
disagree with our arguments 
and are convinced that FSI enhance the QCD penguin contribution to $\epe$ 
bringing it within the SM to agree with the data. Unfortunately no public response in arxiv to our claim in \cite{Buras:2016fys} has been made by these authors
and they were unable to demonstrate in private conversations that we are wrong.
While lattice QCD calculations are likely to resolve this controversy in due 
time, it would be desirable if other experts had a look at \cite{Buras:2016fys}
 in order to confirm our claim or invalidate it explicitly in public.

But the controversy on FSI is not the whole story. Also the existence of meson 
evolution, crucial in the DQCD approach, has been questioned over the last 30 
years by some chiral and lattice experts. We have demonstrated in \cite{Buras:2018lgu} that meson evolution is crucial in reproducing the results for four
BSM $K^0-\bar K^0$ matrix elements from lattice QCD  (ETM, SWME and RBC-UKQCD) collaborations \cite{Carrasco:2015pra,Jang:2015sla,Boyle:2017ssm}.
The important 
virtue of this exercise is the absence of FSI so that the meson evolution 
can be better exposed than in $K\to\pi\pi$ decays.
In this analysis it is also demonstrated that in contrast to $\hat B_K$, 
the strict large $N$ limit  in the case of the BSM parameters $B_{2-5}$ is 
a very bad approximation missing lattice results by  factors of $2-3$. 
Including 
non-factorizable contributions represented by meson evolution allows to
understand the lattice data. 

 The latter analysis demonstrates the importance of the QCD dynamics at scales 
below $1\gev$ and gives additional support to our claim that meson evolution 
 is the dominant QCD dynamics responsible for the $\Delta I=1/2$ rule and also 
 the $\epe$ anomaly. We are not aware of any analytical approach that could provide such 
insight in lattice QCD results in question. We challenge the chiral perturbation theory experts to provide an insight into the values of $B_i$ parameters from Lattice QCD in their framework, in particular without using lower energy constants obtained from lattice QCD.  Motivated by this result we now turn our attention to NP.

\section{Master Formula for $\epe$ beyond the SM}\label{sec:5}
Having both the RG evolution and all matrix elements at the low-energy scale
$\muLow$ for the first time at hand allowed us recently \cite{Aebischer:2018quc}
to present a master formula for  $(\epe)_\text{BSM}$ that exhibits
its dependence on each Wilson coefficient at the scale $\muEW$ and consequently
is valid in {\em any} theory beyond the SM that is free from non-standard light degrees
of freedom below the electroweak scale. We will now  present 
this formula. Technical details which led to this formula can be found 
in  \cite{Aebischer:2018quc} and in particular in \cite{Aebischer:2018csl}.

Writing $\epe$ as a sum of the SM and BSM contributions,
\begin{align}
  \frac{\varepsilon'}{\varepsilon} &
  = \left(\frac{\varepsilon'}{\varepsilon}\right)_\text{SM}
  + \left(\frac{\varepsilon'}{\varepsilon}\right)_\text{BSM} \,,
\end{align}
the master formula of \cite{Aebischer:2018quc} for the BSM part then reads
\begin{align}
  \label{eq:master}
  \left(\frac{\varepsilon'}{\varepsilon}\right)_\text{BSM} &
  = \sum_i  P_i(\muEW) ~\text{Im}\left[ C_i(\muEW) - C^\prime_i(\muEW)\right]
  \times (1\,\text{TeV})^2,
\end{align}
where
\begin{align}
  \label{eq:master2}
  P_i(\muEW) & = \sum_{j} \sum_{I=0,2} p_{ij}^{(I)}(\muEW, \muLow)
  \,\left[\frac{\langle O_j (\muLow) \rangle_I}{\text{GeV}^3}\right],
\end{align}
with the sum over $i$ extending over the Wilson coefficients $C_i$ of all
operators and their chirality-flipped counterparts, that is $36 + 36'$
linearly independent four-quark operators and $1 + 1'$ chromo-magnetic dipole operators.
The $C_i'$ are the Wilson coefficients of the corresponding
chirality-flipped operators obtained by interchanging $P_L\leftrightarrow P_R$.  The
relative minus sign accounts for the fact that their $K\to\pi\pi$ matrix
elements differ by a sign.  Among the contributing operators are also operators
present already in the SM but their Wilson coefficients in (\ref{eq:master})
include only BSM contributions. In view of space limitations we cannot list
all these operators here and the readers are invited to look up 
our last three papers \cite{Aebischer:2018rrz,Aebischer:2018quc,Aebischer:2018csl}.

The dimensionless coefficients $p_{ij}^{(I)}(\muEW,\muLow)$ include the QCD and
QED RG evolution from $\muEW$ to $\muLow$ for each
Wilson coefficient as well as the relative suppression of the contributions to
the $I=0$ amplitude due to ${\text{Re}A_2} / {\text{Re}A_0}\ll 1$ for the matrix
elements $\langle O_j (\muLow) \rangle_I$ of all the operators $O_j$ present at
the low-energy scale. The index $j$ includes also $i$ so that the effect of
self-mixing is included. The $P_i(\muEW)$ do not depend on $\muLow$ to the
considered order, because the $\muLow$-dependence cancels between matrix
elements and the RG evolution operator.  Moreover, it should be emphasized that
their values are {\em model-independent} and depend only on the SM dynamics
below the electroweak scale, which includes short distance contributions down to
$\muLow$ and the long distance contributions represented by the hadronic matrix
elements. The BSM dependence enters our master formula in (\ref{eq:master}) {\em
  only} through the Wilson coefficients $C_i(\muEW)$ and
$C^\prime_i(\muEW)$. That is, even if a given $P_i$ is non-zero, the fate of its
contribution depends on the difference of these two coefficients. In particular,
in models with exact left-right symmetry this contribution vanishes as first
pointed out in \cite{Branco:1982wp}.

The numerical values of the $P_i(\muEW)$ are collected in the tables
presented in \cite{Aebischer:2018quc,Aebischer:2018csl}.
 As seen in (\ref{eq:master2}),
the $P_i$ depend on the hadronic matrix elements
$\langle O_j (\muLow) \rangle_I$ and the RG evolution factors
$p_{ij}^{(I)}(\muEW, \muLow)$.  The numerical values of the hadronic matrix
elements rely on lattice QCD in the case of SM operators  and DQCD in the case
of BSM operators as summarized above.

Inspecting the results in the tables in \cite{Aebischer:2018csl} the following
observations can be made, here given in a simplified form in 
terms of three islands of hadronic matrix elements: SM-like island governed by 
SM operators $Q_{7,8}$, chromo-magnetic one governed by chromo-magnetic penguin 
operator $O_{8g}$ and  the BSM-island conquered in \cite{Aebischer:2018rrz}:
\begin{itemize}
\item The largest $P_i$ values on SM-like island can be traced back to the large values
  of the matrix elements $\langle Q_{7,8}\rangle_2$, the dominant electroweak
  penguin operators in the SM, and the enhancement by $22$ of the
  $I=2$ contributions relative to $I=0$ ones.
\item The small $P_i$ values on the chromo-magnetic island  are the consequence of the fact that
  each one is proportional to $\langle O_{8g} \rangle_0$, which has recently been
  found to be much smaller than previously expected \cite{Constantinou:2017sgv,
    Buras:2018evv}. Moreover, as $\langle O_{8g}\rangle_2=0$, all contributions
  in this class are suppressed by the factor $22$ relative to
  contributions from other classes.
\item The large $P_i$ values on the BSM-island can be traced back to the large
  hadronic matrix elements of scalar and tensor operators calculated recently in
  \cite{Aebischer:2018rrz}. 
\end{itemize}

The question then arises which creatures are living on these islands. Here comes
the answer.

\section{Lessons from the SMEFT Analysis of $\epe$}\label{sec:6}
In \cite{Aebischer:2018csl} we have presented
 for the first time a model-independent anatomy of the ratio $\epe$
in the context of  the $\Delta S = 1$ EFT with operators invariant
under QCD and QED and in the context of the Standard
Model Effective Field Theory (SMEFT) with the operators invariant under the full SM gauge group. This was only possible thanks to the very recent
 calculations of the $K\to\pi\pi$ matrix elements of BSM operators,
namely of the chromo-magnetic dipole operators by lattice QCD
\cite{Constantinou:2017sgv} and DQCD \cite{Buras:2018evv} and in particular through the calculation of matrix elements of all four-quark BSM operators, including scalar and tensor operators, by DQCD \cite{Aebischer:2018rrz}. Even if the
latter calculations have been performed in the chiral limit, they offer for the
first time a look into the world of BSM operators contributing to $\epe$.

Our main goal was to identify those NP scenarios which
are probed by $\epe$
and which could help to
explain the emerging anomaly in $\epe$ discussed above.

In the last three years a number of analyses, addressing the $\epe$ anomaly
in concrete models, appeared in the literature. They are collected in Table~1
of \cite{Buras:2018wmb} and also listed in  \cite{Aebischer:2018csl}.
But all these analyses concentrated on   models in which NP entered exclusively
through modifications of the Wilson coefficients of SM operators. In particular the Wilson coefficient of the dominant electroweak penguin operator $Q_8$. Thus 
these analyses have been performed on the SM-like island.
This
is a significant limitation if one wants to have a general view of possible BSM scenarios
responsible for the $\epe$ anomaly. In particular, in the absence of even approximate
values of hadronic matrix elements of BSM operators, no complete model-independent analysis was possible until recently.

The recent calculations of BSM $K\to\pi\pi$ matrix elements, in particular
of those of scalar and tensor operators in \cite{Aebischer:2018rrz} allowed 
to perform the analysis also on the BSM-island which 
combined
with the EFT and in particular SMEFT technology in \cite{Aebischer:2018csl} widened
significantly our view on BSM contributions to $\epe$. The analysis in \cite{Aebischer:2018csl} has two main virtues:
\begin{itemize}
\item
It opens the road  to the analyses of $\epe$ in any theory beyond the
SM and allows with the help of the master formula  in (\ref{eq:master})
to search very efficiently
for BSM scenarios behind the $\epe$ anomaly. In particular the values
of $P_i$ collected in  these two papers indicate which routes could
be more successful than  others both in the context of the low-energy
EFT and SMEFT.
By implementing our results in the open source code \texttt{flavio} \cite{flavio},
testing specific BSM theories becomes particularly simple.
\item
Through our SMEFT analysis we were able to identify
correlations between $\epe$ and various observables that depend sensitively
on the operators involved. Here $\Delta S=2$, $\Delta C=2$ and electro-magnetic
dipole moments (EDM)
play a prominent role but also correlations with $\Delta S=1$ and $\Delta C=1$
provide valuable informations. We ask the interested readers to look at our
papers.
\end{itemize}

Our main messages to take home are as follows:
\begin{itemize}
\item
Tree-level vector exchanges, like $Z^\prime$, $W^\prime$ and $G^\prime$ contributions (SM-like island), discussed
already by various authors and vector-like quarks \cite{Bobeth:2016llm} can be responsible for the observed anomaly but
generally one has to face important constraints from  $\Delta S=2$ and $\Delta C=2$
transitions as well as direct searches and often some fine tuning is required. Here
the main role is played by the electroweak penguin operator $Q_8$ with its Wilson coefficient
significantly modified by NP.
\item
Models with tree-level exchanges of heavy colourless or coloured scalars (BSM-island)  are a new avenue,
opened with the results for BSM operators from DQCD in \cite{Aebischer:2018rrz}.
In particular scalar and tensor operators, having chirally enhanced matrix elements
and consequently large coefficients $P_i$, are candidates for the explanation
of the anomaly in question. Moreover, some of these models, in contrast to
models with  tree-level $Z^\prime$ and $G^\prime$ exchanges,
are free
from  both  $\Delta S=2$ and $\Delta C=2$ constraints.
The EDM of the neutron is an important constraint for these models,
depending on the
couplings, but does not preclude a sizable NP effect in $\epe$.
\item
Models with modified $W^\pm$ or $Z^0$ couplings can induce sizable
effects in $\epe$ without appreciable constraints from semi-leptonic
decays such as $\kpn$ or $\klpll$.
In the case of a SM singlet $Z'$ mixing with the $Z^0$, sizable $Z^0$-mediated contributions are disfavoured by electroweak precision tests. Yet, as discussed
in \cite{Buras:2018wmb} and in references given there, also such models could
 play some role in understanding the role of NP in $\epe$. This is in particular  the case of model with vector-like quarks \cite{Bobeth:2016llm}.
\end{itemize}

The future of $\epe$ in the SM and in the context of searches for NP will depend
on how accurately it can be calculated. This requires improved lattice calculations
not only of the matrix elements of SM operators but also of the BSM ones, which
are known presently only from the DQCD approach in the chiral limit. Moreover, 
the impact of FSI on the values of $P_i$ have to be investigated. As these 
values are dominated by $I=2$ matrix elements we expect that these effects amount to at most $10-20\%$ of the present values. It is also
hoped that lattice QCD will be able to take into account  isospin breaking corrections
and that other lattice collaborations will attempt to calculate
hadronic matrix elements of all relevant operators. In this context we hope
that the new analysis of the RBC-UKQCD collaboration with improved matrix elements
to be expected this year will shed new light on the hinted anomaly. Such future updates
can be easily accounted for by the supplementary details on the master formula in the appendices of \cite{Aebischer:2018csl}.

On the short-distance side the NNLO results for QCD penguins should be available soon
\cite{Cerda-Sevilla:2016yzo}. The dominant NNLO corrections to electroweak penguins
have been calculated almost 20 years ago \cite{Buras:1999st}. 
As pointed out in \cite{Aebischer:2018csl}
all the existing estimates of $\epe$ at
NLO suffer from short-distance renormalization scheme uncertainties
in the electroweak penguin contributions and also scale uncertainties in 
$m_t(\mu)$ that are removed only in the NNLO matching at the electroweak scale \cite{Buras:1999st}. In the naive dimensional regularization
(NDR) scheme, used in all recent analyses, these corrections enhance the
electroweak penguin contribution by roughly $16\%$, thereby leading to a {\em negative}
shift of $-1.3\times 10^{-4}$ decreasing further the value of $\epe$, similarly to
isospin breaking effects.
This could appear small in
view of other uncertainties. However, on the one hand, potential scale and
renormalization scheme uncertainties have been removed in this manner and on
the other hand, one day such corrections could turn out to be relevant. Finally,
the fact that this correction further decreases $\epe$ within the SM gives
another motivation for the search for new physics responsible
for it, and thus for our recent  analyses. 
 
As far as BSM operators are concerned,
a NLO analysis of their Wilson coefficients is in progress, but its importance
is not as high as of hadronic matrix elements due to significant additional
parametric uncertainties residing  in any  NP model.
In any case, in the coming years the ratio $\epe$ is expected to play a significant
role in the search for NP. In this respect, the results presented in
\cite{Aebischer:2018quc,Aebischer:2018csl}  will be
helpful in disentangling potential models of new CP violating sources beyond the
SM as well as constraining the magnitude of their effects.

\section{Outlook}\label{outlook}\label{sec:7}
Our outlook is very short. The future of $\epe$, in particular in correlation 
with rare decays $\kpn$ and $\klpn$, looks great and the coming years should be very exciting. I am looking forward to QCD@Work 2020 
when the landscape of NP will be more transparent than it is now.

%
%

\section*{Acknowledgements}
 It is 
a pleasure to thank the organizers of this workshop for inviting me 
to this interesting event and for an impressive hospitality. Particular thanks 
go to my four great collaborators: Jason Aebischer, Christoph Bobeth,  Jean-Marc
 G{\'e}rard and David Straub for fantastic time we spent together in the last six months. This research was  
supported by the DFG cluster of excellence ``Origin and Structure of the Universe''.

\bibstyle{woc}
\bibliography{allrefs}
\end{document}